\documentclass[12pt]{article}
\usepackage{amsmath}
\usepackage{bm}
\usepackage{color}

\begin{document}
\begin{center}
\begin{large}
{\bf Macroscopic body in Snyder space and minimal length estimation}
\end{large}
\end{center}

\centerline {Kh. P. Gnatenko \footnote{E-Mail address: khrystyna.gnatenko@gmail.com}, V. M. Tkachuk \footnote{E-Mail address: voltkachuk@gmail.com}}
\medskip
\centerline {\small \it Ivan Franko National University of Lviv, Department for Theoretical Physics,}
\centerline {\small \it 12 Drahomanov St., Lviv, 79005, Ukraine}

\abstract{We study a problem of description of macroscopic body motion in the frame of nonrelativistic Snyder model. It is found that the motion of the center-of-mass of a body is described by an effective parameter which depends on the parameters of Snyder algebra for coordinates and momenta of particles forming the body and their masses. We also show that there is reduction of the effective parameter with respect to parameters of Snyder algebra for coordinates and momenta of individual particles. As a result the problem of extremely small result for the minimal length obtained on the basis of studies of the Mercury motion in the Snyder space is solved. In addition we find that relation of parameter of Snyder algebra with mass opens possibility to preserve the property of independence of the kinetic energy on composition, to recover the weak equivalence principle, to consider coordinates as kinematic variables, to recover proportionality of momenta to mass and to consider Snyder algebra for coordinates and momenta of the center-of-mass of a body defined in the traditional way.
}

\section{Introduction}

Idea of noncommutativity of coordinates was proposed by Heisenberg for solving the problem of ultraviolet divergences in quantum field theory. Formalization of this idea was done by Snyder in 1947 \cite{Snyder}. Development of String Theory and Quantum Gravity (see, for instance, \cite{Witten,Doplicher}) leads to recent growing of interest to studies of space with noncommutative structure.

The Snyder model is based on the following commutation relations for coordinates and momenta
\begin{eqnarray}
[X_{\mu},X_{\mu}]=i\hbar\beta^2J_{\mu\nu},{}\label{0al}\\{}
[X_i,P_j]=i\hbar(\eta_{\mu\nu}+\beta^2 P_{\mu}P_{\nu}),{}\label{0al1}\\{}
[P_{\mu},P_{\nu}]=0,\label{0al2}
 \end{eqnarray}
where $J_{\mu\nu}$ are the Lorentz generators, $\eta_{\mu\nu}$ is the metric tensor $[\eta_{\mu\nu}]=\textrm{diag}(-1,1,1,1)$, $\beta^2$ is a constant. Algebra (\ref{0al})-(\ref{0al2}) is invariant under the Lorentz transformations and leads to the minimal length (see, for instance, \cite{Mignemi11}).

Different problems were examined in the frame of the Snyder algebra. Among them, for example, are area quantization  \cite{RomeroPLB}, hamiltonian formalism \cite{Lu}, free particle  \cite{Mignemi11,Mignemi12}, harmonic oscillator \cite{Mignemi11,Mignemi12,Leiva10},  planetary motion \cite{Benczik,Leiva,Ivetic,Mignemi14},  and many others.
Also symmetries in Snyder space \cite{Banerjee06,Banerjee11}  algebraic structure of the Snyder space and its physical predictions \cite{Battisti} were examined.

 Studies of many-particle problems opens possibility to find new effects of Snyder algebra on the properties of physical systems and to estimate the minimal length. However, making calculations of the upper bound on the minimal length on the basis of studies of the perihelion shift of the Mercury planet the authors of papers \cite{Ivetic,Mignemi14} faced a problem of extremely small result for the minimal length which is much beyond the Planck length. The problem of macroscopic body in Snyder space is similar to that which appears in doubly spatial relativity and is known as a "soccer-ball problem" \cite{Hossenfelder,Amelino-Camelia,Hinterleitner,HossenfelderS}.

In the present paper we show that the problem of extremely small minimal length obtained in \cite{Ivetic} is caused by the assumption that parameter $\beta$ of the Snyder algebra is the same for elementary particles and macroscopic bodies. We find that the motion of the center-of-mass of macroscopic body is described by an effective parameter which is less than parameter $\beta$ corresponding to individual particles and reexamine the minimal length obtained in \cite{Ivetic} to more relevant one.
Also we propose condition on the parameter $\beta$ on which the kinetic energy of a body is independent on composition, the algebra for coordinates and momenta of the center-of-mass reproduces the Snyder algebra for coordinates and momenta of individual particles and the weak equivalence principle is preserved in the Snyder space.

We would like to note that the problem of violation of the weak equivalence principle appears also in the frame of deformed algebra with minimal length \cite{Tk1}, noncommutative algebra of canonical type \cite{Bastos1,GnatenkoPLA13,Saha,Bertolami2,Saha1,GnatenkoPLA17,GnatenkoEPL}, noncommutative algebra of Lie-type \cite{GnatenkoPRD} and can be solved due to idea to relate parameters of algebra with mass \cite{Tk1,GnatenkoPLA13,GnatenkoPLA17,GnatenkoEPL,GnatenkoPRD}.

The paper is organized as follows. In Section 2 we study features of description of motion of a body in the frame of nonrelativistic Snyder model.  Motion of a particle (a body) in gravitational field in Snyder space and the weak equivalence principle are studied in Section 3.  In Section 4 the upper bound on the minimal length  is examined on the basis of studies of the Mercury motion. Conclusions are presented in Section 5.

\section{Features of description of motion of a body in the frame of nonrelativistic Snyder model}

Let us consider the motion of a body of mass $M$ in the frame of nonrelativistic Snyder model
\begin{eqnarray}
 [X_i,X_j]=i\hbar\beta^2J_{ij},{}\label{1al}\\{}
  [X_i,P_j]=i\hbar(\delta_{ij}+\beta^2P_i P_j),{}\label{11al1}\\{}
  [P_i,P_j]=0,\label{1al2}
 \end{eqnarray}
here $J_{ij}=X_iP_j-X_jP_i$, $i,j=(1,2,3)$ (see, for instance, \cite{Mignemi11}). In the classical limit from
(\ref{1al})-(\ref{1al2}) one obtains the following Poisson brackets
\begin{eqnarray}
 \{X_i,X_j\}=\beta^2J_{ij},{}\label{al}\\{}
  \{X_i,P_j\}=\delta_{ij}+\beta^2P_i P_j,{}\label{al1}\\{}
  \{P_i,P_j\}=0.\label{al2}
 \end{eqnarray}

For a body of mass $M$ with Hamiltonian
\begin{eqnarray}
H=\frac{P^2}{2M},\label{h}
\end{eqnarray}
(here $P^2=\sum_iP_i^2$) taking into account relations of Snyder algebra (\ref{al})-(\ref{al2}), we find
\begin{eqnarray}
\dot{X}_i=\{X_i,H\}=\frac{P_i}{M}(1+\beta^2P^2),\label{eq1}\\
\dot{P}_i=0
\end{eqnarray}
Using (\ref{eq1}), up to the first order in $\beta^2$  Hamiltonian (\ref{h}) can be rewritten as a function of velocity of the body
\begin{eqnarray}
H=\frac{M\dot{X}^2}{2}(1-2\beta^2M^2\dot{X}^2)\label{hh}
\end{eqnarray}
here $X^2=\sum_iX_i^2$.

Let us study the case when the body can be divided into $N$ parts with masses $m_a$ which can be considered as particles. The particles are tightly bound and move with the same velocities as the body. So, on the other hand according to the additivity property of the kinetic energy one has
\begin{eqnarray}
H=\sum_a\frac{(P^{(a)})^2}{2m_a}\label{hadd}
\end{eqnarray}
where index $a$ label the particles.
 For coordinates $X^{(a)}_i$ and momenta $P^{(a)}_i$ of particles in Snyder space in general case relations (\ref{al})-(\ref{al2}) can be written as
\begin{eqnarray}
\{X^{(a)}_i,X^{(b)}_j\}=\delta_{ab}\beta_a^2J^{(a)}_{ij},{}\label{nt}\\{}
\{X^{(a)}_i,P^{(b)}_j\}=\delta_{ab}(\delta_{ij}+\beta_a^2P^{(a)}_i P^{(a)}_j),{}\label{nt0}\\{}
\{P^{(a)}_i,P^{(b)}_j\}=0.\label{nt1}
\end{eqnarray}
Here $J^{(a)}_{ij}=X^{(a)}_iP^{(a)}_j-X^{(a)}_jP^{(a)}_i$, indexes $a$, $b$ label the particles. Writing (\ref{nt})-(\ref{nt1}), we assume that Poisson brackets for coordinates and momenta of different particles vanish and study a general case when parameters $\beta_a$ are different for different particles.
Using (\ref{nt})-(\ref{nt1}), we find
\begin{eqnarray}
\dot{X}^{(a)}_i=\frac{P^{(a)}_i}{m_a}(1+\beta_a^2(P^{(a)})^2),\label{eqp1}
\end{eqnarray}
and up to the first order in $\beta_a^2$ expression for  kinetic energy (\ref{hadd})  as a function of velocity reads
\begin{eqnarray}
H=\sum_a\frac{m_a\dot{X}^2}{2}(1-2\beta_a^2m_a^2\dot{X}^2)=\nonumber\\=\frac{M\dot{X}^2}{2}(1-2M^2\sum_a\beta_a^2\mu_a^3\dot{X}^2).\label{hadd1}
\end{eqnarray}
(here we take into account that the velocities of particles are the same and equal to the velocity of the body $\dot{X}^{(a)}_i=\dot{X}_i$). Comparing expressions (\ref{hh}), (\ref{hadd1}) one has that parameter of Snyder algebra $\beta$, corresponding to the body, is defined as
\begin{eqnarray}
\beta^2=\tilde\beta^2=\sum_a\beta_a^2\mu^3_a \label{beff}.
\end{eqnarray}
We would like to stress here that if we consider parameter of Snyder algebra to be the same for different particles, for different bodies, namely if $\beta=\beta_a$,  from (\ref{hh}), (\ref{hadd1}) one has that the fundamental property of the kinetic energy, its additivity, is violated.

Another fundamental property of the kinetic energy is its independence of composition.
Note that the kinetic energy (\ref{hadd1}) depends on the effective parameter (\ref{beff}) which is determined by the masses of particles forming the body and parameters $\beta_a$. So, the kinetic energy of a body in Snyder space depends on its composition.
It is important to note that if we consider the idea to relate parameter of Snyder algebra with mass, namely if we suppose that the relation
\begin{eqnarray}
\beta_a m_a=\gamma=const,\label{c1}
\end{eqnarray}
is satisfied (here $m_a$ is the mass of a particle, $\gamma$ is a constant which is the same for different particles) from (\ref{beff}) one obtains
\begin{eqnarray}
\tilde\beta=\frac{\gamma}{M}\label{beffc}
\end{eqnarray}
and the kinetic energy (\ref{hadd1}) reads
\begin{eqnarray}
H=\frac{M\dot{X}^2}{2}(1-2\gamma^2\dot{X}^2).\label{hadd2}
\end{eqnarray}
 So, due to condition (\ref{c1}) the kinetic energy depends on the total mass of the system and do not depend on its composition. So, the property of independence of the kinetic energy of composition is recovered in Snyder space.

We would like to note here that the relation of parameter of Snyder algebra with mass was also considered in \cite{Banerjee06}. The authors of paper \cite{Banerjee06} proposed the form of an action yielding the Snyder algebra with parameter proportional inversely to mass.
It is also worth mentioning that the idea to relate parameters of deformed algebra with mass opens possibility to recover independence of kinetic energy on composition in deformed space with minimal length \cite{Tk1}, in noncommutative space of canonical type \cite{GnatenkoPLA13}, in noncommutative phase space \cite{GnatenkoPLA17}.

In addition due to assumption (\ref{c1}) Poisson brackets for coordinates and momenta of the center-of-mass of a body defined in traditional way
\begin{eqnarray}
\tilde{\bf X}=\sum_a\mu_a{\bf X}^{(a)},\label{totalc}\\
\tilde{\bf P}=\sum_a{\bf P}^{(a)},\label{totalm}
 \end{eqnarray}
 reproduce relations of Snyder algebra (\ref{al})-(\ref{al2}) with effective parameter (\ref{beffc}).

Taking into account (\ref{nt})-(\ref{nt1}), for coordinates and momenta of the center-of-mass (\ref{totalc}), (\ref{totalm}) one obtains
\begin{eqnarray}
\{\tilde{X}_i,\tilde{X}_j\}=\sum_a\mu_a^2\beta_a^2J^{(a)}_{ij}\label{cmc}\\
\{\tilde{X}_i,\tilde{P}_j\}=\delta_{ij}+\sum_a\mu_a\beta_a^2P^{(a)}_{i}P^{(a)}_{j}\label{cmc1}\\
\{\tilde{P}_i,\tilde{P}_j\}=0.
\end{eqnarray}
Note that relations (\ref{cmc}), (\ref{cmc1}) do not reproduce relations of Snyder algebra (\ref{al}), (\ref{al1}).

If condition (\ref{c1}) holds, using (\ref{eqp1}) one can write
\begin{eqnarray}
\frac{P^{(a)}_i}{m_a}\left(1+\gamma^2\left(\frac{P^{(a)}}{m_a}\right)^2\right)=\dot{X}^{(a)}_i\label{eqp11}
\end{eqnarray}
From (\ref{eqp11}) one has that ratio ${P^{(a)}_i}/{m_a}$ is determined by a constant $\gamma$ and velocity $\dot{X}^{(a)}_i$. So,  in Snyder space  for particles which move with the same velocities  one has
\begin{eqnarray}
\frac{P^{(a)}_i}{m_a}=\frac{\tilde{P}}{M}.\label{ratio}
\end{eqnarray}
Taking into account (\ref{c1}), (\ref{cmc}), (\ref{cmc1}), (\ref{ratio}) for coordinates and momenta of the center-of-mass of a body one obtains relations of Snyder algebra with parameter $\tilde{\beta}^2$ given by (\ref{beffc}). Namely one has
 \begin{eqnarray}
\{\tilde{X}_i,\tilde{X}_j\}=\tilde{\beta}^2\tilde{J}_{ij},\label{22}\\
\{\tilde{X}_i,\tilde{P}_j\}=\delta_{ij}+\tilde{\beta}^2\tilde{P}_{i}\tilde{P}_{j},\label{222}
\end{eqnarray}
where
\begin{eqnarray}
\tilde{J}_{ij}=\tilde{X}_i\tilde{P}_j-\tilde{X}_j\tilde{P}_i,\label{tj}
\end{eqnarray}
Note that expressions (\ref{22}), (\ref{222})   are obtained without making approximations connecting with smallness of the value $\beta^2$. In all orders in $\beta^2$ due to condition (\ref{c1}), the relations for coordinates and momenta of a body reproduce relations of Snyder algebra with $\tilde{\beta}^2$.

It is important that the parameter $\tilde \beta$ corresponding to the motion of the center-of-mass of a body is less than parameters $\beta_a$ corresponding to the motion of particles forming it. For instance if a body (a system) consists of $N$ particles with masses $m$ and parameters $\beta$ according to (\ref{beff})  one has that
\begin{eqnarray}
\tilde\beta^2=\frac{\beta^2}{N^2}.\label{reduction}
\end{eqnarray}
The same result can be derived from (\ref{c1}) and (\ref{beffc}). So, there is reduction of effective parameter $\tilde \beta^2$ with respect to $\beta^2$ due to multiplier $1/N^2$. Therefore, from (\ref{reduction}) follows that  effect of features of space structures on the Planck scale on the macroscopic systems is less than this effect on elementary particles. This statement is naturally understandable and should be taken into account considering macroscopic bodies in Snyder space.  On the basis of this conclusion in Section 4  we explain and reexamine extremely small result for estimation of the minimal length obtained on the basis of studies of the Mercury motion in the frame of Snyder algebra.

At the end of this section we would like to note another important result which can be obtained due to relation (\ref{c1}).
Coordinates and momenta which satisfy (\ref{nt})-(\ref{nt1}) can be represented as
\begin{eqnarray}
X^{(a)}_i={x^{(a)}_i}{\sqrt{1-\beta_a^2({p}^{(a)})^2}},\label{rep1}\\
P^{(a)}_i=\frac{p^{(a)}_i}{\sqrt{1-\beta_a^2({p}^{(a)})^2}},\label{rep2}
\end{eqnarray}
with $({p}^{(a)})^2=\sum_i({p}^{(a)}_i)^2$. Coordinates and momenta $x^{(a)}_i$, $p^{(a)}_i$ satisfy the ordinary relations
\begin{eqnarray}
\{x^{(a)}_i,x^{(b)}_j\}=0,\label{or}\\
\{x^{(a)}_i,p^{(b)}_j\}=\delta_{ij}\delta_{ab},\\
\{p^{(a)}_i,p^{(b)}_j\}=0.\label{or1}
\end{eqnarray}
Momenta are bounded as $(p^{(a)})^2<1/\beta_a^2$.
From (\ref{rep1}) follows that coordinates depend on momenta and therefore depend on the mass.
 Also, from (\ref{rep2}) one has that the momenta are not proportional to mass because expression under the square root depends on the squared momenta. If condition (\ref{c1}) holds one can write
\begin{eqnarray}
X^{(a)}_i={x^{(a)}_i}{\sqrt{1-\gamma^2({p}^{(a)})^2/m_a^2}},\label{crep1}\\
P^{(a)}_i=\frac{p^{(a)}_i}{\sqrt{1-\gamma^2({p}^{(a)})^2/m_a^2}}.\label{crep2}
\end{eqnarray}
So, one has that the coordinates (\ref{crep1}) do not depend on mass and can be considered as kinematic variables, also momenta  (\ref{crep2}) are proportional to mass as it should be, if we consider parameters $\beta_a$ to be proportional inversely to mass (\ref{c1}).

\section{Motion in gravitational field in Snyder space and the weak equivalence principle}

Let us study the motion of a particle of mass $m$ in the gravitational field in Snyder space.
The Hamiltonian corresponding to the particle reads
\begin{eqnarray}
H=\frac{{P}^2}{2m}+mV(X_1,X_2,X_3).\label{hh12}
\end{eqnarray}
Coordinates $X_i$ and momenta $P_i$ satisfy (\ref{al})-(\ref{al2}), function $V(X_1,X_2,X_3)$ describes the field. So, one can write the following equations of motion
\begin{eqnarray}
\dot X_i=\frac{P_i}{m}\left(1+\beta^2P^2\right)+m\beta^2J_{ij}\frac{\partial V}{\partial X_j},\label{vel}\\
\dot P_i=-m\frac{\partial V}{\partial X_i}-m\beta^2P_i P_j\frac{\partial V}{\partial X_j}.
\end{eqnarray}
We would like to note that because of relations (\ref{al}), (\ref{al1}) the velocity of a particle in the gravitational field depends on its mass (\ref{vel}). It is important to stress that if relation (\ref{c1}) holds one has
\begin{eqnarray}
\dot X_i={P^{\prime}_i}\left(1+\gamma^2(P^{\prime})^2\right)+\gamma^2(X_iP^{\prime}_j-X_jP^{\prime}_i)\frac{\partial V}{\partial X_j},\label{v1}\\
\dot P^{\prime}_i=-\frac{\partial V}{\partial X_i}-\gamma^2P^{\prime}_i P^{\prime}_j\frac{\partial V}{\partial X_j},\label{v2}
\end{eqnarray}
where we denote $P^{\prime}_i=P_i/m$. Equations (\ref{v1}) and (\ref{v2}) do not depend on mass therefore their solutions $X_i(t)$, $P^{\prime}_i(t)$ do not depend on mass too. So, the weak equivalence principle is preserved in Snyder space if parameter $\beta$ is determined by mass as (\ref{c1}).

In more general case of motion of a body of mass $M$ in gravitational field
if condition (\ref{c1}) is satisfied, writing Hamiltonian of the body in the form
 \begin{eqnarray}
H=\frac{{\tilde{P}}^2}{2M}+MV({\tilde X}_1,{\tilde X}_2,{\tilde X}_3),\label{hh12}
\end{eqnarray}
and taking into account  (\ref{22}), (\ref{222}),  one obtains equations of motion for the center-of-mass as (\ref{v1}), (\ref{v2}) with effective parameter (\ref{beffc}) which do not depend on the mass of the body and on its composition. So, the weak equivalence principle is recovered due to relation (\ref{c1}).

At the end of this section we would like to note that in \cite{Mignemi14} the authors examined orbit of a particle in Schwarzschild space-time in the frame of Snyder model. It was obtained that the equivalence principle is not preserved because of terms proportional to $\beta^2m^2$ in the corrections to the geodesics motion. Note that if relation (\ref{c1}) is satisfied one has $\beta^2m^2=\gamma^2$. So, the corrections depends on constant $\gamma$ which does not depend on mass and due to condition (\ref{c1}) the equivalence principle is recovered in Schwarzschild space-time.

\section{Estimation of the upper bound on the minimal length on the basis of studies of the Mercury motion}

The motion of a particle in the Coulomb potential in the frame of relations (\ref{al})-(\ref{al2}) was studied in \cite{Benczik,Ivetic,Leiva}.
The authors of papers examined the Kepler problem in the frame of Snyder space, applied their result to the case of Mercury motion and found the perihelion shift
\begin{eqnarray}
\delta\theta=-\frac{2\pi\beta^2Gm^2M}{a(1-e^2)}\label{est1}
\end{eqnarray}
caused by  deformation (here $G$ is the gravitational constant, $M$ is the mass of the Sun, $m$ is the mass of Mercury, $e$ is the eccentricity, $a$ is the semi-major axis).  On the basis of comparison of this result with discrepancy from the prediction of general relativity of the perihelion precession of Mercury which is of the order $10^{-12}{\textrm rad/rev}$ very strong (well below the Planck scale)  restriction on the value of $\beta$  was found \cite{Ivetic}. Such paradoxical effect was obtained because of assumption that the motion of  Mercury planet is described by the same parameter $\beta$ as the motion of a particle. As we have shown in Section 2 the motion of  macroscopic body in Snyder space is described by effective parameter $\tilde \beta$.  Therefore  in (\ref{est1}) $\beta$ should be replaced by $\tilde\beta$ defined as (\ref{beffc}).

According to (\ref{c1}), (\ref{beffc}) we can relate parameter of Snyder algebra $\tilde{\beta}$ corresponding to the Mercury planet with parameter $\beta^2_{nuc}$ which describes motion of nucleons in the Snyder space as
\begin{eqnarray}
\tilde{\beta}^2=\frac{\beta^2_{nuc}m^2_{nuc}}{m^2},\label{effM1}
\end{eqnarray}
where $m_{nuc}$ is the mass of nucleon.
Taking into account (\ref{effM1}) and assuming that
\begin{eqnarray}
\frac{2\pi\tilde{\beta}^2Gm^2M}{a(1-e^2)}<10^{-12},
\end{eqnarray}
as was done in \cite{Ivetic}, for the minimal length we have
 \begin{eqnarray}
\hbar\beta_{nuc}<10^{-19}m.\label{est12}
\end{eqnarray}
This result is $16$ orders above the Planck length therefore is more relevant one and is in agreement to that obtained for the minimal length for nucleons in deformed space \cite{Tk2}.

\section{Conclusion}

Features of description of motion of a body in the Snyder space have been examined considering a general case when coordinates and momenta of different particles satisfy Snyder algebra with different parameters  (\ref{nt})-(\ref{nt1}).
We have shown that the motion of a body in the Snyder space is described by effective parameter $\tilde{\beta}$ (\ref{beff}).
It is important to conclude that the parameter $\tilde{\beta}$ is less than parameters $\beta_a$ corresponding to particles forming the body. In particular case of a system of $N$ particles with masses $m$ and parameters $\beta$ one has $\tilde{\beta}=\beta/N$ (see (\ref{reduction})). Therefore, we have concluded that effect of space quantization on the macroscopic systems is less than on elementary particles. This conclusion is naturally understandable and should be taken into account in studies of macroscopic bodies in the frame of the Snyder model.

We have shown that the fundamental property of the kinetic energy, its additivity, is violated if one assume that coordinates and momenta of a particle and coordinates and momenta  of the center-of-mass of a body satisfy relations of the Snyder algebra with the same parameters $\beta$. Besides such assumption leads to paradoxical effect, namely an extremely small result for the minimal length \cite{Ivetic}.  We have found that  the result for the minimal length \cite{Ivetic} can be reexamined to relevant one if we take into account that the motion of the center-of-mass of a body is described by an effective parameter. We have estimated the upper bound for the minimal length  $10^{-19}m$ (\ref{est12}) which is above the Planck length and  is in agreement to that obtained in deformed space \cite{Tk2}.

We have also proposed condition on the parameter $\beta$ of Snyder algebra which relates it with mass (\ref{c1}) and opens possibility to solve number of problems in Snyder space. Namely, we have shown that if parameter $\beta$ is inversely proportional to mass the kinetic energy is independent on the composition, the  weak equivalence principle is recovered, the coordinates can be considered as kinematic variables and momenta are proportional to mass as it should be. Besides we have found that on the condition (\ref{c1}) the Poisson brackets for coordinates and momenta of the center-of-mass of a body reproduce relations of Snyder algebra with effective parameter $\tilde{\beta}$ (\ref{22}), (\ref{222}).

The number of results which can be obtained due to idea to relate parameter of Snyder algebra with mass justifies  the importance of the idea. In addition we would like to note that relation of parameters of deformed algebra with mass is important for solving number problems in deformed space with minimal length \cite{Tk1,Tk2,Tk3}, noncommutative space of canonical type \cite{GnatenkoPLA13,GnatenkoPLA17,GnatenkoEPL}, in a space with noncommutativity of Lie-type \cite{GnatenkoPRD}.

\section*{Acknowledgements}

This work was partly supported by the Project $\Phi\Phi$-63Hp
(No. 0117U007190) from the Ministry of Education
and Science of Ukraine.


\begin{thebibliography}{99}
\bibitem{Snyder} H.~Snyder, Phys. Rev. {\bf71}, 38 (1947).
\bibitem{Witten} N. Seiberg, E. Witten, J. High Energy Phys. {\bf 9909}, 032 (1999).
\bibitem{Doplicher} S. Doplicher, K. Fredenhagen, J.E. Roberts, Phys. Lett. B {\bf 331}, 39 (1994).
\bibitem{Mignemi11} S. Mignemi, Phys. Rev. D {\bf84}, 025021 (2011).

\bibitem{RomeroPLB} J. Romero, A. Zamora, Phys. Lett. B {\bf 661}, 11 (2008).
\bibitem{Lu} L. Lu, A. Stern,  Nucl. Phys. B {\bf 860}, 186 (2012).
\bibitem{Mignemi12} S Mignemi, Class. Quantum Grav. {\bf29}, 215019 (2012).
\bibitem{Leiva10} C. Leiva, Pramana J. Phys.  {\bf 74},  169 (2010).

\bibitem{Benczik} S. Benczik, L.N. Chang, D. Minic, N. Okamura, S. Rayyan, T. Takeuchi, Phys. Rev. D {\bf66}, 026003 (2002).
\bibitem{Leiva} C. Leiva, J. Saavedra, J. R. Villanueva, Pramana J. Phys. {\bf80}, 945 (2013).
\bibitem{Ivetic} B. Ivetic, S. Meljanac, S. Mignemi, Class. Quantum Grav. {\bf31}, 105010 (2014).
\bibitem{Mignemi14} S. Mignemi, R. Strajn,  Phys. Rev. D {\bf90}, 044019 (2014).
\bibitem{Banerjee06} R. Banerjee, S. Kulkarni, S.  Samanta JHEP 05 (2006) 077.
\bibitem{Banerjee11} R. Banerjee, K. Kumar, D. Roychowdhury, JHEP 03 (2011) 060.
\bibitem{Battisti} 	M. V. Battisti, S. Meljanac, Phys. Rev. D {\bf 79} 067505 (2009).
\bibitem{Hossenfelder} S. Hossenfelder Phys. Rev. D {\bf75}, 105005 (2007).
\bibitem{Amelino-Camelia}  G. Amelino-Camelia, Symmetry {\bf2}  230 (2010).
\bibitem{Hinterleitner}  F. Hinterleitner, Phys. Rev. D {\bf71}, 025016 (2005).
\bibitem{HossenfelderS} S. Hossenfelder, SIGMA {\bf 10}, 074 (2014).
\bibitem{GnatenkoPLA17} Kh. P. Gnatenko, V. M. Tkachuk, Phys. Lett. A {\bf381}, 2463 (2017).
\bibitem{Bastos1} C. Bastos, O. Bertolami, N. C. Dias, J. N. Prata, Class. Quant. Grav. {\bf 28}, 125007 (2011).
\bibitem{GnatenkoPLA13}  Kh.P.~Gnatenko,  {Phys. Lett. A} {\bf 377}, 3061 (2013).
\bibitem{Saha} A. Saha, Phys. Rev. D. {\bf89}, 025010 (2014).
\bibitem{Bertolami2} O. Bertolami, P. Leal, Phys. Lett. B  {\bf750}, 6 (2015).
\bibitem{Saha1} S. Bhattacharyya, S. Gangopadhyay, A. Saha, EPL {\bf120}, 30005 (2018).
\bibitem{GnatenkoEPL} Kh. P. Gnatenko, EPL  {\bf 123}, 50002 (2018).
\bibitem{GnatenkoPRD} Kh. P. Gnatenko, Phys. Rev. D {\bf 99}, 026009 (2019).
\bibitem{Tk1}  V. M. Tkachuk, Phys. Rev. A {\bf86}, 062112 (2012).
\bibitem{Tk2}  C. Quesne, V.M. Tkachuk, Phys. Rev. A {\bf81}, 012106 (2010).
\bibitem{Tk3} V. M. Tkachuk, Found. Phys. {\bf46}, 1666 (2016).
\end{thebibliography}
\end{document}